 \documentclass[preprint2]{aastex}
\usepackage {graphicx}

\newcommand{\timesM}{$\times$}
\newcommand{\Msun}{{\rm M}_{\solar}}
\newcommand{\msun}{${\rm M}_{\odot}$}
\newcommand{\solar}{\ifmmode_{\mathord\odot}\;\else$_{\mathord\odot}\;$\fi}
\newcommand{\cm}{{\rm cm}^{-3}}
\newcommand{\ergs}{{\rm erg} \thinspace  {\rm s}^{-1}}
\newcommand{\yr}{\rm\thinspace yr^{-1}}
\newcommand{\kpcTwo}{\rm\thinspace Kpc^{-2}}
\newcommand{\pcTwo}{\rm\thinspace pc^{-2}}

\shorttitle{Stellar Feedback and Galaxy Formation}
\shortauthors{Ceverino & Klypin}

\begin{document}

\title{The role of stellar feedback in the formation of galaxies.}

\author{Daniel Ceverino and Anatoly Klypin}
\affil{Astronomy Department, New Mexico State University, Las Cruces, NM}


\begin{abstract}
Although supernova explosions and stellar winds happen at very small
scales, they affect the interstellar medium (ISM) at galactic scales
and regulate the formation of a whole galaxy.  Previous attempts of
mimicking these effects in simulations of galaxy formation use very
simplified assumptions. We develop a much more realistic prescription
for modeling the feedback, which minimizes any ad hoc sub-grid
physics.  We start with developing high resolution models of the ISM
and formulate the conditions required for its realistic functionality:
formation of multi-phase medium with hot chimneys, super-bubbles, cold
molecular phase, and very slow consumption of gas. We find that this
can be achieved only by doing what the real Universe does: formation
of dense ($> 10$ H atoms cm$^{-3}$), cold ($T\approx 100$ K) molecular
phase, where the star formation happens, and which is disrupted by
young stars.   Another important
ingredient is the runaway stars: massive binary stars ejected from
molecular clouds when one of the companions becomes a supernova. Those
stars can move to 10-100 parsecs away from molecular clouds before
exploding themselves as supernovae. This greatly facilitates the
feedback. Once those effects are implemented into cosmological
simulations, galaxy formation proceeds more realistically. For
example, we do not have the overcooling problem. The angular momentum
problem (resulting in a too massive bulge) is also reduced
substantially: the rotation curves are nearly flat.  The galaxy
formation also becomes more violent. Just as often observed in QSO
absorption lines, there are substantial outflows from forming and
active galaxies. At high redshifts we routinely find gas with few
hundred km s$^{-1}$ and occasionally  $1000-2000$ km s$^{-1}.$
The gas has high metallicity, which may exceed the solar
metallicity. The temperature of the gas in the outflows and in
chimneys can be very high: $T=10^7-10^8$ K.  The density profile of
dark matter is still consistent with a cuspy profile.  The simulations
reproduce this picture only if the resolution is very high: better
than 50 pc, which is 10 times better than the typical resolution in
previous cosmological simulations. Our simulations of galaxy formation
reach the resolution of 35 pc.
\end{abstract}


\keywords{hydrodynamics, methods: n-body simulations, ISM: general, stars: formation,  galaxies: formation, evolution}

\section{Introduction}

The current cosmological paradigm, the $\Lambda$CDM Universe, has
successfully explained the overall assembly of cosmic structures
\citep{Blumenthal84, Davis85, Spergel07}.  In this picture ordinary
matter (``baryons''), which emits and absorbs light, passively follows
the evolution of the dark matter. This should be corrected, if we want
to make a realistic theory of galaxy formation. It is necessary to
include the physics of the gas and galaxy formation into the
$\Lambda$CDM paradigm, because, after all, many observational
evidences of cosmic structures come from the light emitted by
galaxies.

Galaxy formation is driven by a complex set of physical processes with
very different spatial scales. Radiative cooling, star formation and
supernova explosions happen at scales less than 1~pc, but they affect the
formation of a whole galaxy \citep{Dekel86}. In addition,
large-scale cosmological processes, such as gas accretion through
cosmic filaments, and galaxy mergers, control the galaxy assembly. As
a result, a complex interplay between very different processes drives
the formation of galaxies.  
Cosmological gasdynamical simulations have
become useful tools to study galaxy formation.

In early cosmological simulations ``galaxies'' formed with too small
disks and a significant fraction of angular momentum was lost
\citep{Navarro00}. The situation has improved in the last
years. \citet{Sommer03,Governato04, Kaufmann06} show that a
sub-kpc resolution is necessary to prevent an artificial loss of
angular momentum.  Recent improvements in both the resolution
and modeling of the feedback have
resulted in simulations with extended galactic disks 
\citep{Governato04, Robertson04, Brook04, Okamoto05,
Elena06, Governato07}. However,
simulated galaxies are still too concentrated, and more realistic
simulations with better resolution and better physics are needed to
reproduce the shape of the rotation curves of observed galaxies.

Current simulations lack the necessary resolution to follow correctly
the effect of supernova explosions in the ISM. Because of this lack of
resolution, the modeling of stellar feedback has relied on ad-hoc
assumptions about the effect of stellar feedback at scales unresolved
by simulations (0.2-1 Kpc). Early attempts to introduce stellar
feedback into simulations found the obstacle of a strong radiative
cooling. The energy deposited by supernova explosions was quickly
radiated away without any effect in the ISM \citep{Katz92}. Several
shortcuts have been proposed to get around this over-cooling problem.

The most common method is to artificially stop cooling when the
stellar energy is deposited
\citep{Gerritsen97,Thacker00,Sommer03,Keres05,Governato07} This approach
prolongs the adiabatic phase of supernova explosion (the Sedov
solution) to about 30 Myr. The motivation behind this ad-hoc
assumption is that the combination of blastwaves from different
supernova explosions and turbulent motions produces hot bubbles much
larger than individual supernova remnants and last longer. All this
effects are not resolved with the current resolution. They do not
develop in a self-consistent way. Instead, the delay in the cooling is
introduced by hand. The problem is that other effects can be missed at
the same time due to a lack of resolution and an inaccurate modeling
of feedback.

Another method is to introduce a sub-resolution model in which the
energy from supernova explosions is stored in an unresolved hot phase,
which does not cool and looses energy through the evaporation of
cold clouds \citep{Yepes97,Springel03}.
 In this model, the only effect of stellar feedback is to regulate the star
formation: the hot gas is coupled with the cold
phase through cloud formation and evaporation. As a result, this high
entropy gas is artificially trapped within the galactic disk. Thus,
galactic winds are introduced in a simplified way in order to reproduce other
natural effects of stellar feedback, such as galactic outflows.

An alternative approach assumes kinetic feedback instead of thermal
feedback \citep{Navarro93}. In that case, the energy from
supernova explosions or stellar winds is transfered to the kinetic
energy of the surrounding medium. This energy is not dissipated
directly by radiative cooling. However, in order to resolve this
effect accurately, simulations should be able to resolve the expansion
of individual supernova explosions or the stellar winds from individual
stars. Currently, this is not possible. At larger scales, the picture
is more complicated. Different blastwaves from different supernova
explosions can collide, dissipating their kinetic energy. The same
dissipation of energy happens in collisions of stellar winds in
stellar clusters. So, it is commonly assumed that most of the kinetic energy
from stellar feedback is dissipated into thermal energy at the
smallest scales resolved by simulations. Nevertheless, this
feedback-heated gas can expand. As a result, thermal energy can be
transfered to kinetic energy. The net results are flows at large
scales powered by the thermal feedback. However, feedback heating should
dominates over radiative cooling: only in this case those flows are
produced.

To summarize, the main problems of current simulations of galaxy
formation are the lack of the necessary resolution and too simplified
models of the complex hydrodynamic processes in the multiphase ISM.

The galactic ISM has a very wide range of densities and temperatures
(for review see \citet{Cox05} and \citet{Ferriere01}). Three distinct
phases are distinguished: the dense cold gas (giant molecular clouds
(GMC), cold HI gas or diffuse clouds) with densities above 10
cm$^{-3}$ and temperatures bellow 100 K, the warm component with
densities between 0.1 and 1 cm$^{-3}$ and temperatures of several
thousands degrees, and the hot phase with temperatures above $10^5$ K
and densities bellow $10^{-2}$ cm$^{-3}$. This multiphase medium is
set by the competition of cooling and heating mechanism and the onset
of thermal instabilities.  The hot ISM component $(T>10^5$ K) is
usually associated with gas heated by shocks. They can be produced by
turbulent motions driven by gravitational and thermal
instabilities. However, these turbulent driven shocks can only heat
the gas up to $10^6$~K \citep{Wada01}.  Only supernova explosions and
stellar winds can produce larger gas temperatures \citep{Mccray79,
Spitzer90}.

2D and 3D hydrodynamical simulations of the ISM have enough resolution
(parsecs) to resolve the multi-phase nature of the ISM and to explore
complicated effects of stellar feedback on different scales
\citep{Rosen95,Scalo98,Korpi99,deAvillez00,deAvillez04, deAvillez07,
Wada01, Wada07,Slyz05}.  There is much to learn from
these simulations.  However, they typically focus on conditions in the
solar neighborhood, which are different from what one may expect
during early stages of galaxy formation.  Not always they follow the
whole gas cycle: cooling, star formation, and stellar feedback.  For
example, \citet{deAvillez04} include star formation but artificially
restrict the rate of supernova explosions around a fixed
value. However, this rate could be much higher in large star forming
regions. As a result, the effect of stellar feedback is underestimated
in these regions. Nevertheless, the effect of the stellar feedback in
the ISM, such as the formation of hot bubbles and super-bubbles is
resolved.

It is crucial to understand where and how the energy from massive
stars is released back to the ISM. While a large fraction of massive
stars are found in stellar clusters and OB associations, 10-30\% are
found in the field, away from any molecular cloud or stellar cluster
\citep{Gies87, Stone91}. This population have peculiar
kinematics. Their velocity dispersion is about 30~km~s$^{-1}$, much
higher than the velocity dispersion of the population of massive stars
in clusters (10 km s$^{-1}$) \citep{Stone91}.  Some of these stars
have large peculiar velocities, up to 200 km s$^{-1}$
\citep{Hoo00}. This is why they are called runaway stars.
 The current scenario of the origin of runaway stars is the ejection
of these massive stars from stellar clusters.  There are two possible
mechanisms of this ejection.  One possibility is the ejection due to a
supernova explosion in a close binary system \citep{Zwicky57, Blaauw61}.
The second mechanism is the ejection due to dynamical
encounters in the crowded regions of stellar clusters \citep{Poveda67}. 
 In spite of the fact that a significant fraction of the stellar
feedback occurs far  from star forming regions, no attention
has been paid to its effect on the galaxy formation.

We first study the effect of stellar feedback in the ISM, using
simulations of a Kpc-scale piece of the ISM with few parsecs
resolution. Then, we check if this picture holds when the resolution
is degraded to the resolution that our cosmological simulations can
achieve at high redshift. Finally, we study the effect of stellar
feedback in galaxy formation at high redshift using cosmological
hydrodynamical simulations.

This paper is organized as follows. Section 2 describes the necessary
conditions in which stellar feedback dominates over radiative
cooling. Section 3 describes all details of the modeling of stellar
feedback.  Section 4 shows Kpc-scale simulations of the ISM. Section 5
describes the cosmological simulations of galaxy formation. Finally,
section 6 is the discussion and conclusion. Throughout the paper we
give quantities in physical units.


\section{Physical conditions for the heating regime}

The thermodynamical state of the gas depends on two competing
processes: heating from stellar feedback and cooling from radiative
processes. They appear as source and sink terms of internal energy
in the equation of the first law of thermodynamics:
\begin{equation}
\frac{du}{dt} + p \nabla \cdot \mathbf{v} =   \Gamma - \Lambda
\label{eq:1}
\end{equation}
where $u$ is the internal energy per unit volume, $p$ is the pressure of
the gas, and \textbf{v} is its velocity. Parameter  $\Gamma$ is the
heating rate due to stellar feedback, and $\Lambda$ is the net cooling
rate from radiative processes.

The heating rate from stellar feedback can be expressed as the rate of
energy losses from a young and active single stellar population with a
given density, $\rho_{*,\rm young}:$
\begin{equation}
\Gamma  =  \rho_{*,\rm young}  \Gamma'
\label{eq:2}
\end{equation}
where  $\Gamma'$ is the specific rate of energy losses of the stellar
population according to its age.
The cooling rate can be expressed as:
\begin{equation}
 \Lambda  =  n^2_H \Lambda',
\label{eq:3}
\end{equation}
\noindent where $n_H$ is the hydrogen number density.

\subsection{Heating versus radiative cooling}

Now, we can ask ourselves under which conditions the feedback heating
dominates over the radiative looses. Using the expression, $n_H=\rho_{\rm gas}/
(\mu_H m_H),$ where $\rho_{\rm gas}$ is the gas density, $\mu_H$ is the
molecular weight per hydrogen atom and $m_H$ is the hydrogen mass, the
condition for heating $( \Lambda \leq \Gamma )$ can be expressed as:
\begin{equation}
n_H \Lambda'  \leq  \frac{\rho_{*,\rm young}}{\rho_{\rm gas}} \mu_H m_H \Gamma'
\label{eq:4}
\end{equation}
Using typical values, we can rewrite the condition for the
heating regime in the following way:
\begin{equation}
 \left( \frac{n_H}{0.1  \thinspace \cm} \right) \left( \frac{\Lambda'}{10^{-22} \thinspace \ergs \thinspace  \cm} \right)   \leq
\label{eq:4b}
\end{equation}
\begin{eqnarray*}
 \left(\frac{\rho_{*,\rm young}}{\rho_{\rm gas}}\right) \left( \frac{\Gamma'}{10^{34} \thinspace \ergs \thinspace \Msun^{-1}} \right)
\end{eqnarray*}
The cooling rate, $\Lambda',$ is a strong function of gas
temperature. So, the temperature and the density of the gas are two
key properties in establishing the cooling or the heating regimes. The
following two examples illustrate common situations.

At temperatures around $10^4$~K, the cooling rate is close to its
maximum value. We use $\Lambda'= 10^{-22} $ erg s$^{-1}$ cm$^3$ as a
fiducial value. In this case eq.(\ref{eq:4b}) shows that the heating
overcomes the cooling only at very low densities $n_H \leq 0.1 $
$\cm,$ optimistically assuming that the ratio of densities,
$\rho_{*,\rm young}/\rho_{\rm gas}$ is about unity. As a result,
stellar feedback is not able to heat the gas beyond $10^4$ K for
densities higher than 0.1~$\cm$ and typical values of $\Gamma'.$ This
is the well known overcooling problem for simulations, which allow
cooling only to a temperature of $10^4$~K at which the star formation
is assumed to happen. The energy from stellar feedback is radiated
away very efficiently and the thermal feedback cannot play any
role. In this case one needs to invoke ``subgrid physics'' -- a guess
how the system should react to the energy released by the stars.

The situation is completely different if the gas is allowed to cool to
100 K. The cooling is very inefficient at that temperature: $\Lambda'=
10^{-25}$ erg s$^{-1}$ cm$^3$. So, stellar feedback can produce the
net gas heating even if the density is large: $n_H \approx 100 $ $\cm$
for $\rho_{*,\rm young} \approx \rho_{\rm gas}$. Our conclusion is
that simulations should include cooling process bellow $10^4$ K. The
cold phase should be resolved in order to get a high efficiency of
stellar feedback.


However, heating to high temperatures is still problematic because as
the gas is heated, the cooling rate increases.  So, the peak of the
cooling rate at $10^4$~K is a bottle-neck for heating gas to higher
temperatures.
Nevertheless, temperatures of diffuse gas as high as $10^6-10^7$~K
have been observed around star-forming regions such as the Rosette
nebulae \citep{Townsley03, Wang07}, M17 \citep{Townsley03}, and the
Orion nebula \citep{Feigelson05,Guedel07}.  The main question is how
young and massive stars can heat their surrounding medium to these
high temperatures, if the original medium, in which they were born had
high densities.

The answer likely depends on the distance from those young stellar
clusters.  At small 1-2~pc distances it is likely to have the collisions of
stellar winds \citep{Townsley03,Feigelson05} . At larger distances the
heating is related with the formation of superbubbles: the cumulative
effect of winds and shocks generated by many young stars. One way or
another, the density of gas around the young stellar population
decreases and the ratio $\rho_{*,\rm young}/\rho_{\rm gas}$ increases
as the over-pressured bubble of gas expands. Once the density goes
below $0.1$ $\cm$, eq.(\ref{eq:4b}) can be fulfilled even at $10^4$
K. The net result is a heating regime, in which the surrounding gas
can be heated to very high temperatures.  In other words, the process
starts with the expanding bubbles at low temperatures and then
proceeds to a runaway overheating regime.
 
As an example, we consider a typical GMC with a mass
of $10^5$~\msun~ and a size of 50~pc.  These are the typical values
found in recent catalogs of GMCs in M33
\citep{Rosolowsky07}, M31 and the Milky way \citep{Sheth08}.
Therefore, the mean density is $n_H=50 $ $\cm$. This value seems low
compared with typical observed densities of molecular clouds. However,
GMCs are highly clumpy. High-density clumps are
usually embedded in a low density inter-clump medium. As a result, the
volume-averaged density inside clouds is much smaller than the typical
observed mass-weighted density \citep{McKee99}.

Now, we consider an Orion-like stellar cluster formed at the center of
the cloud. The mass of the stellar cluster is $ 5 \times 10^3
\thinspace \Msun$ \citep{Hillenbrand98}. In a region of mass
$10^4$~\msun, the stellar cluster has the ratio $\rho_{*,\rm
{young}}/\rho_{\rm gas}$ equal to 0.5, and the condition for heating,
eq.(\ref{eq:4b}), is fulfilled. This heating produces an over-pressured
hot bubble with a pressure 100 times higher than the surrounding
unperturbed medium. As a result, the bubble expands, the density
decreases, and the ratio $\rho_{*,\rm young}/\rho_{\rm gas}$
increases. Then, we get a runaway bubble, which proceeds to blowing
away all gas \citep{Kroupa01}.

Simulations should resolve the expansion of bubbles over-pressured by
stellar feedback.  The density of young (and active) stars and the
density of gas should be comparable at the smallest scales resolved by
the simulations.  The minimum value of the ratio $\rho_{*,\rm
young}/\rho_{\rm gas}$ depends on the gas density
(eq.~\ref{eq:4b}). For moderate gas densities, $n_H= 10 -100 $ $\cm,$
the above ratio should be around 0.1-1.

The above condition can be achieved if the star formation efficiency,
the fraction of the progenitor cloud consumed in stars is 10\%-50\% at
the resolution scale. This high efficiency is consistent with the
observed value of 10\%-40\% found in Galactic stellar clusters,
\citep{GreeneYoung92, Elmegreen00, Kroupa01}. Due to the fact that
80\% of the Galactic star formation occurs in stellar clusters
\citep{LadaLada03}, this high efficiency of star formation should be
considered in any star formation model which can resolve the sites where
star formation occurs.

\subsection{Local gravity versus pressure gradient}

As we saw in the previous section, low densities are required in order
to heat the gas beyond the peak of the cooling curve. Stellar feedback
should evacuate the gas by creating an expanding bubble around young
stellar clusters.  However, the over-pressured bubble expands only if
the pressure gradient overcomes self-gravity.

If we consider an over-pressured bubble of radius $R$ in a homogeneous
medium of density $\rho$, we can derive a Jeans-instability type of
condition. As a result, the bubble expands only if the difference in
pressure with its surroundings, $\Delta P$, satisfies the following
relationship:
\begin{equation}
\Delta P/k  \geq  \frac{4 \pi}{3k} G (\rho R)^2 = 10^{-1}
(n_H R_{pc})^2
\label{eq:5}
\end{equation}
where $k$ is the Boltzmann constant, $G$ is the gravitational constant,
and $R_{\rm pc}$ is the radius in pc.  The above equation sets the
conditions for the bubble expansion. For the Galactic plane the
pressure is  $P/k \sim 2 \times 10^4 $ $\cm K$ \citep{Cox05}. For
example, a region of 50~pc in radius and a density of 100~$\cm$ will
only expand, if the difference in pressure is bigger than $ 2 \times
10^6 $ $\cm$~K. This can be achieved, if the bubble is over-pressured
by more than 100 times. Stellar feedback can produce this overpressure
just by raising the temperature from 100~K to $10^4$~K. The resulted
over-pressured region will expand, and the density as well as the
cooling rate will decrease. So, the efficiency of stellar feedback
increases, raising the temperature and pressure further.


Eq. \ref{eq:5} also sets a upper limit on the resolution. Using the
equation of state of the ideal gas $P= n k T$, where $n$ is the mean
number density and T is the temperature of the gas, the over-pressured
bubble should be resolved with a spatial resolution $X_{pc}=
R_{pc}/2$, such that the expansion is resolved:
\begin{equation}
 \left( \frac{X_{pc}}{ 75 \thinspace {\rm pc}} \right)^2 \leq \left( \frac{T}{10^4 \thinspace   {\rm K} }
\right) \left(\frac{n_H}{10 \thinspace  \cm} \right)^{-1} 
\label{eq:5b}
\end{equation} 
As a result, for typical values of these over-pressured regions, the
resolution should be better than $\sim$ 70 pc. Otherwise, the bubble
cannot overcome its self-gravity and cannot expand.

\section{Stellar feedback model}


We assume a model of thermal feedback for the injection of energy from
stellar winds and supernova explosions. The kinetic energy from these
processes is efficiently dissipated into thermal energy due to shocks
at scales bellow the spatial resolution. 

The net thermal rate $(\Gamma - \Lambda )$ is used to update the
internal energy in each step of the simulation. This approach is
rather different than the deposition of energy. Instead, the energy
injection from stellar feedback is treated in a self-consistent way
along with the radiative looses.

\subsection{Heating rate from stellar feedback}

\begin{figure}[tb!]
\epsscale{1.}
\plotone{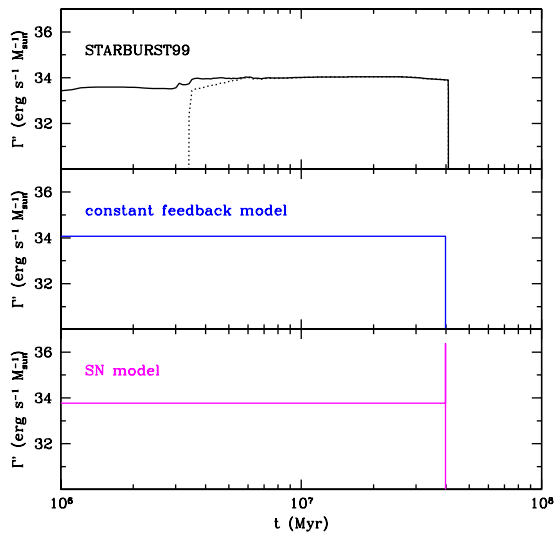}
\caption{Rate of energy losses per unit mass from a single stellar
population. Top panel shows the results from the STARBURST99 code, assuming
a Miller-Scalo IMF for a mass range $(0.1-100)$~\msun. The dotted line
shows the contribution of supernova explosions and the full line shows
the total rate. Although supernova explosions dominate the overall
energy release, stellar winds are the only mechanism of energy release
during the first few Myr. Middle and bottom panels show two different
models: a constant feedback model and a model of stellar wind plus
core-collapse supernova. Although the total energy released is the
same in both models, the SN model is more elementary and takes into
account the explosive nature of core-collapse supernova.
 }
\label{fig:1}
\end{figure}

The heating rate from stellar feedback in a given volume element is
modeled as the rate of energy losses from a set of single stellar
populations present in that volume. This is just a generalization of
eq.(\ref{eq:2}):
\begin{equation}
\Gamma  =  \frac{1}{V} \sum_i M_i \Gamma'(t_i),
\end{equation}
where $M_i$ and $t_i$ are the mass and the age of each single stellar
population.

The modeling of the specific release of energy over time, $\Gamma',$ is
motivated by the results from population synthesis codes, such as
STARBURST99 \citep{Leitherer99}. Figure~\ref{fig:1} shows different models
of $\Gamma'$ and the results of a STARBURST99 computation with a
Miller-Scalo IMF from 0.1~\msun~  to 100~\msun. Parameter $\Gamma'$ is
dominated by stellar winds from massive OB main-sequence stars and WR
stars during the first few Myr. Later  the energy  is
produced by core-collapse supernovae from stars more massive than 
8~\msun. After 40~Myr, the release of energy comes from stellar winds
of AGB stars and other less powerful sources, and the injection rate
drops 6 orders of magnitudes.  Supernovae Ia
dominate the feedback at much longer time-scales. We assume a peak of
the SNIa rate at 1~Gyr. However, this peak is 3 orders of magnitude
lower than the contribution from core-collapse supernovae. This is
because the energy from a population of SNIa is diluted over a much longer time scale
than the energy from core-collapse supernovae.

We model $\Gamma'$ with a constant rate of 1.18 \timesM 10$^{34}$ erg
s$^{-1} \Msun^{-1}$ over 40~Myr. This is equivalent to the injection
of $2\times 10^{51}$~erg of energy from stellar winds and supernova
explosions per each massive star with $M> 8$~\msun~ during its lifetime. We
assume a Miller-Scalo IMF in the mass range  $(0.1-100)$~\msun.
Note that  this constant heating rate is the sum of the
contributions from all massive stars in a single stellar
population. We also consider a more simple model, which we call a SN model.
In this case  $10^{51}$~ergs is injected at constant rate due to stellar
winds  over 10 or 40~Myr. Then it follows  a strong peak of energy release  due
to the supernova explosion, in which $10^{51}$~erg are released during
$10^5$~yrs -- the typical age of young supernova remnants.
Although the total energy released is the same in both models, the
SN model takes into account the explosive nature of core-collapse
supernovae.

\subsection{A model of runaway stars}

The effect of runaway stars is implemented by adding a random velocity
to a fraction of stellar particles (10\%-30\%). This extra velocity
has a random orientation and the value is taken from an exponential
distribution with a characteristic scale of 17 km s$^{-1}$. This choice is motivated
by Hipparcos data \citep{Hoo00} and Monte-Carlo simulations
\citep{Dray05}. For comparison, a Gaussian distribution is also
used \citep{Stone91}. However, the effect of runaway stars in the ISM is
not very sensitive to the details of this velocity distribution.

\subsection{Radiative cooling}

Radiative cooling counterbalances feedback heating. So it is very
important to have an accurate model of radiative cooling in order to
study the net effect of stellar feedback in the ISM.

We use the model of radiative cooling described in
\citet{Kravtsov}. It is a metallicity-dependent cooling plus a UV
heating due to a cosmological ionizing background
\citep{Haardt96}. The model includes Compton heating/cooling and
molecular cooling. The temperature range of the model is between
$10^2$ K $< T < 10^9$ K. Thus, this model includes cooling below
$10^4$ K and the gas can reach the thermodynamical conditions of
molecular clouds. As we saw in section 2, this is crucial for the
efficiency of the stellar feedback.

The cooling and heating rates from radiative processes are tabulated
using the CLOUDY code (version 96b4; Ferland et al. 1998). As a result,
the net cooling rate from radiative processes, $\Lambda',$ is available
for a given density, temperature, metallicity and redshift.


\subsection{Description of the code}

The numerical simulations were performed using the Eulerian
gasdynamics + N-body Adaptive Refinement Tree code \citep{Kravtsov97,
Kravtsov99, Kravtsov}. The physical processes of the gas include star
formation, stellar feedback, metal enrichment, self-consistent
advection of metals, cooling and heating rates from
metallicity-dependent cooling and UV heating due to a cosmological
ionizing background.

\section{Results of ISM runs}

Our first step in the understanding of stellar feedback in galaxies is
to understand its effect in the ISM at galactic scales. Therefore, we
run simulations of a $4 \times 4 \times 4$ Kpc$^{3}$ piece of a
galactic disk with 8-16 pc resolution. These simulations fully resolve
the effect of massive stars at galactic scales.  So, resolution is not
longer an issue.

We can use this ISM-scale simulation as a benchmark for the effect of
stellar feedback at galactic scales. Then, we can degrade the
resolution to see which model of feedback reproduces the same overall
picture at lower resolution.  These simulations can then be used as
testing grounds for these models at different resolutions.  They tell
us what are the necessary ingredients to reproduce the truly
effect of stellar feedback at the resolution that we can afford in
cosmological simulations of galaxy formation.

We want to see the effect of stellar feedback in the typical
conditions of normal disk galaxies with moderate gas surface
densities. So, we are not modeling starburst galaxies with large
amounts of gas and high star formation rates. This type of study will
be done in the future.
 
A 4 Kpc box of ISM represents a significant piece of a galactic
disk. The simulation resolves the dense galactic plane, where
molecular clouds are formed. This is important to follow star
formation correctly. At the same time, the simulation follows the gas
at few Kpc above the galactic plane. This height is similar to the
scale-height of the diffuse warm phase of the ISM (Cox 2005).

The simulation includes radiative cooling and UV heating from a
uniform UV field at redshift 0 as described in section 3. Star
formation happens in the highest density peaks with a density
threshold of 100~$\cm$. In each star formation event, 5~\% of the mass
in gas inside a volume element is converted into a stellar particle
with a mass of 88~\msun within a time-step set by the Courant
condition ($\sim$ 2 \timesM 10$^3$ yr).
The supernova model was used for
stellar feedback and SNIa was not included. The metallicity was
assumed solar and constant throughout the simulation.

\begin{figure}[tb!]
\epsscale{0.60}
\plotone{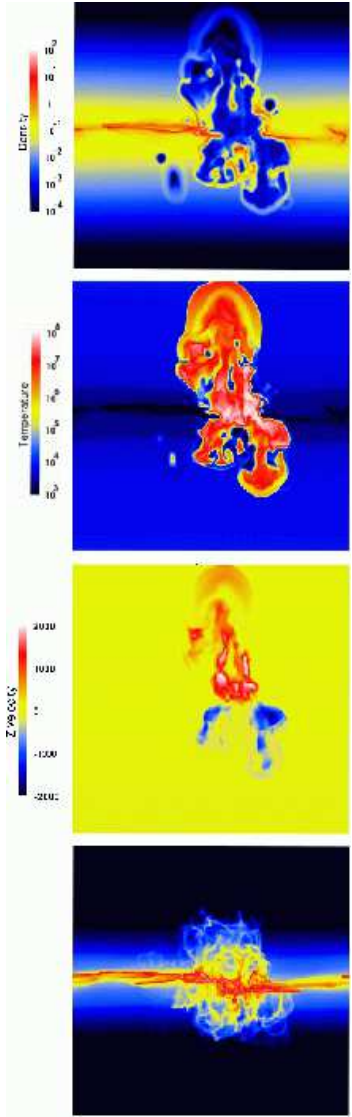}
\caption{Formation of a galactic chimney. Edge-on slices
through the simulation show density, temperature and velocity in the
vertical direction, perpendicular to the galactic plane. 
The bottom panel shows gas column
density. The chimney outflow is not a homogeneous, coherent
flow: it is turbulent and has dense and cold clumps
embedded into the flow. The core of the chimney reaches $10^7$-$10^8$
K. Outflow velocities exceed $10^3$ km~s$^{-1}.$ This hot material
is able to escape the disk and generate a galactic wind.}
\label{fig:2}
\end{figure}

\subsection{Initial conditions}

The initial distribution of gas density is uniform in the $x$ and $y$
directions of the box. In the $z$-direction, the density profile
declines at both sides of the middle plane, $z=z_0=2$ Kpc. This plane defines
the galactic plane for this ISM model:
\begin{eqnarray}
n_H &  = & n_0 \cosh^{-2} \left({\frac{z-z_0}{z_d}}\right)
\end{eqnarray}
where  $n_0$ is the gas density in that plane and $z_d$ is the
scale-height.  

The choice of parameters sets the conditions of a quiescent normal
galactic disk, $n_0= 1 $ $\cm$ and $z_d=250$ pc.  Thus, the surface
density is $,\Sigma_{\rm gas}=16$ $\Msun$ pc$^{-2}.$ The system is
originally in hydrostatic equilibrium with a temperature of $10^4$ K.
No stars are present at the beginning of the simulation.  The box has
open boundaries in the z-direction. So, all material that cross these
boundaries escapes the system.

The initial velocity field consists of a sum of plane-parallel
velocity waves: 
\begin{eqnarray}
u_x  =   \sum_{i,j,k} A_x (i,j,k) \sin (\vec k\cdot\vec r) 
\exp{-  \left(   \frac{z-z_0}{z_d} \right)   ^2} \\
u_y  =   \sum_{i,j,k} A_y (i,j,k)  \sin (\vec k\cdot\vec r)
\exp{-  \left(    \frac{z-z_0}{z_d}   \right)  ^2} \\
u_z  =  \sum_{i,j,k} A_z (i,j,k)  \sin (\vec k\cdot\vec r)
\exp{-   \left(  \frac{z-z_0}{z_d} \right)   ^2} 
\end{eqnarray}
The amplitudes are taken from a Gaussian field with a tilted power
spectrum, $P_k \propto k^{-3},$ where k is the wavenumber,
$k=\frac{2\pi}{L} \sqrt{i^2+j^2+k^2}.$ i,j and k are integers running
from -20 to 20 (excluding 0) and $u_0=20$ km s$^{-1}.$ This is a
typical spectrum of a compressible turbulent medium
\citep{Kraichnan,Vazquez95}.
\begin{eqnarray}
A_x(i,j,k)= u_0 \frac{R_{Gauss}}{(i^2+j^2+k^2)^{3/2}} \\
A_y(i,j,k)= u_0 \frac{R_{Gauss}}{(i^2+j^2+k^2)^{3/2}}  \\
A_z(i,j,k)= u_0 \frac{R_{Gauss}}{(i^2+j^2+k^2)^{3/2}}
\end{eqnarray}
$R_{Gauss}$ is a random number taken from a Gaussian distribution.

\begin{figure}[tb!]
\epsscale{1.0}
\plotone{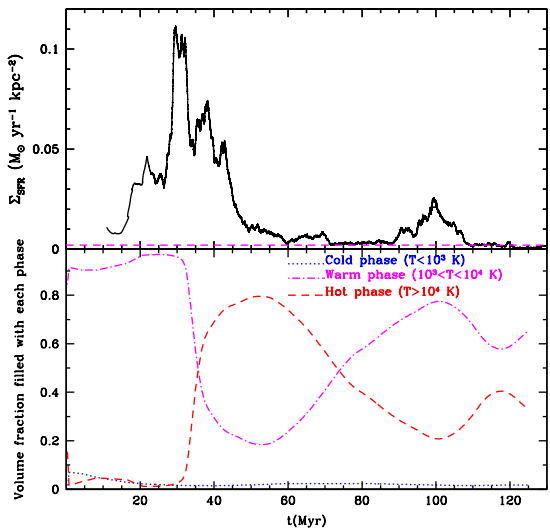}
\caption{Top panel: Star formation rate surface density of the whole
simulation. The value shown is also averaged over a period of $\sim$ 2
\timesM 10$^5$ yr (100 time-steps).  After an initial burst, the star
formation rate surface density is consistent with the
\cite{Kennicutt07} empirical fit (horizontal line).  Bottom panel:
Fraction of volume filled with each gas phase over time.  The volume
occupied by the warm and the hot phase oscillates. The hot phase
dominates after a burst of star formation and the warm phase dominates
when the gas is cooled down.  The cold phase covers a small volume,
which remains constant after an initial collapse.}
\label{fig:3}
\end{figure}

\subsection{Galactic Chimney formation}

At the beginning of the simulation, the gas starts to move according
to the turbulent velocity field. As a result, the gas accumulates
where different flows converge and molecular clouds \footnote{cold and
dense phase with $n_H \geq 30 $ $\cm$ and $T\leq300 K$} naturally
appear in form of filaments and shells. However, around 90\% of the
volume is filled with warm and diffuse gas heated by UV
background. Star formation occurs in the cores of the cold
phase. Newly formed massive stars inject energy and a cavity filled
with hot and very diffuse gas is formed. This over-pressured material
expands and the net result is the formation of super-bubbles. This hot
gas cannot stay in the plane of the disk, as a result, the bubble
expands faster in the direction perpendicular to the disk, because the
density declines in that direction. The bubble develops into a
galactic chimney \citep{Norman89}.  The chimney outflow does not look as a homogeneous,
coherent flow. Instead, the chimney is turbulent and has dense and
cold clumps embedded into the flow.  Eventually, the gas expands in
the halo and cools (Figure \ref{fig:2}).

Another interesting feature seen in this model is a population of
isolated bubbles in the warm medium. These are
the results of individual supernova explosions of runaway stars

\begin{figure*}
\epsscale{1.3}
\plotone{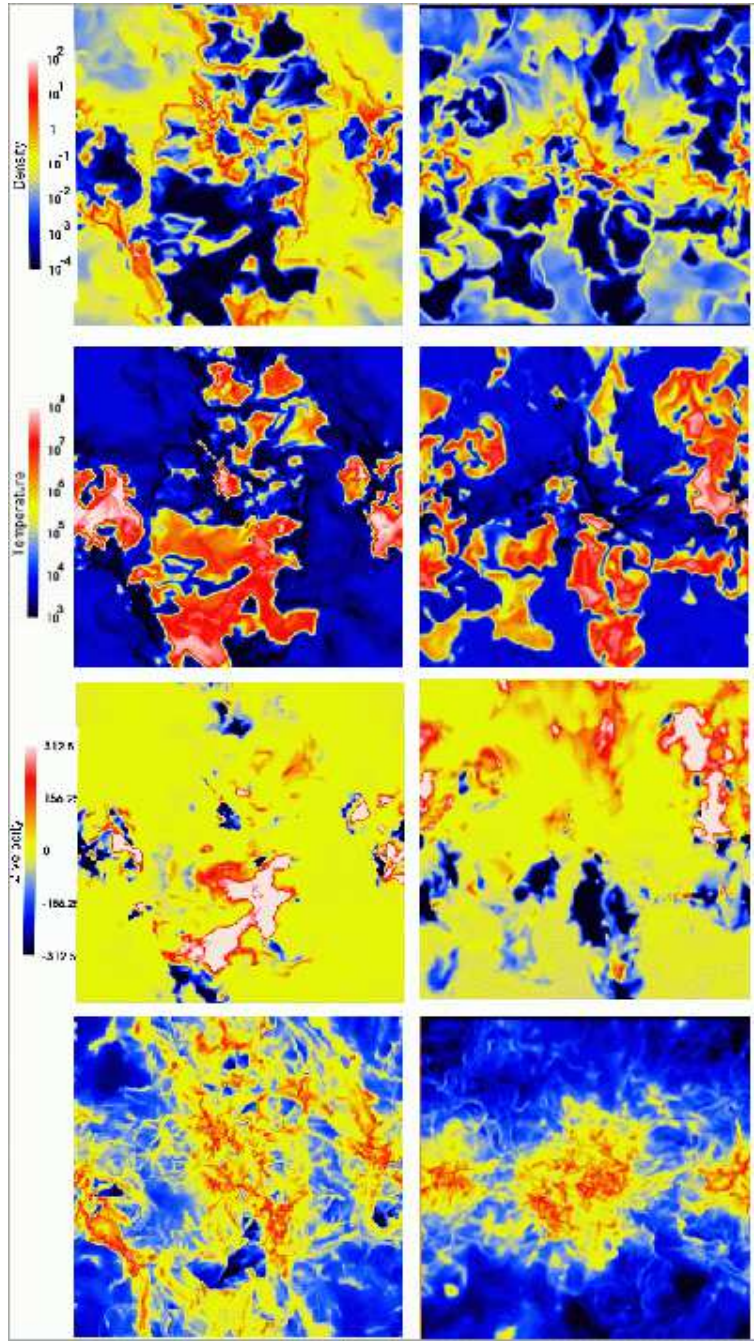}
\caption{ Snapshot of the model after 113 Myr, showing the 
density in $\cm$ (first row), temperature in Kelvin (second row),
gas velocity in the z-direction (third row), and surface density in
$cm^{-2}$ (forth row). Left panels show a face-on view of the galactic
plane $(z=z_0)$ and right panels show an edge-on view perpendicular to
that plane.  The three phases of the ISM are clearly visible: cold and
dense clouds, warm and diffuse medium and hot bubbles with very low
densities.  Velocities exceeding 300 km s$^{-1}$ can be seen in hot
outflows at both sides of the galactic plane.  }
\label{fig:4}
\end{figure*}

\subsection{Star formation rate}

After an initial burst of star formation, the star formation rate is
nearly constant for the rest of the evolution (Figure \ref{fig:3}). We
found a low star formation rate surface density, $\Sigma_{\rm SFR}= 3
\times 10^{-3} \thinspace \Msun \yr \kpcTwo,$ temporally averaged over
a period of $2 \times 10^5$ yr (100 time-steps).  This value is
consistent with the expected value from the correlation between the
star formation rate surface density and the gas surface density found
in nearby galaxies \citep{Kennicutt98,Kennicutt07}. For a gas surface
density of $\Sigma = 12$  $\Msun \pcTwo$ at t=90 Myr, the expected
value from the \citet{Kennicutt07} fit is $\Sigma_{\rm SFR}= 2 \times
10^{-3} \thinspace \Msun \yr \kpcTwo.$ This is very close to our results.

As observers usually do, we also calculate the gas consumption
time-scale, $\tau=$ M$_{\rm GMC}$/SFR, in the simulated molecular
clouds, assuming that gas with a density higher than 30 $\cm$ is
mainly within GMCs. In our simulations, the amount
of gas in molecular clouds is M$_{\rm GMC}$= 8 $\times$ $10^6$ $\Msun$
at t=90 Myr. The star formation rate at that time is ${\rm SFR}=4.8
\times 10^{-2} \thinspace \Msun \yr$. As a result, the gas consumption
time-scale in the simulated clouds is $\tau \approx $ 170 Myr. This is
quite long compared with the typical free-fall time-scale inside
molecular clouds, t$_{\rm ff} = (3\pi/32G\rho)^{1/2} \approx 4$ Myr
for $n_H=$ 100 $\cm.$

In our simulations, the star formation efficiency over a free-fall
time-scale, the fraction of gas consumed in stars during a free-fall
time-scale, is only 2.5\%. This value is consistent with
observations \citep{Zuckerman74}. \citet{KrumholzTa07} report a range
of 0.6\%-2.6\% for the whole population of GMCs of the Milky-Way. Our
value is also consistent with an efficiency of $\sim$ 3\% found in
simulations of GMCs \citep{Vazquez03,Clark05}. Finally, our results
also agree with the model of a turbulent-dominated GMC,
described in \citet{KrumholzMc05}. They give an efficiency per
free-fall time-scale of 1.5\%-3\% for typical values of their model.

After 100 Myr, only 10\% of the gas in the simulation has been
converted into stars.  Our simulations still have plenty of cold ($T
\leq 10^3$ K) gas after 100 Myr. The surface density of this cold gas
is $ \sim 5$ $\Msun \pcTwo$. This value agrees with the surface
density of molecular and atomic hydrogen of $\sim 6$ $\Msun \pcTwo$
found at the solar radius \citep{Ferriere01}. However, the surface
density of molecular gas is low, $ \sim 0.5$ $\Msun \pcTwo$, compared
with the observed value of $ \sim 2.5$ $\Msun \pcTwo$
\citep{Ferriere01}. This partially explains why our star formation
efficiency over a free-fall time-scale is in the higher end of the
observed range.

To conclude, stellar feedback is able to regulate star formation on
galactic scales because it regulates the amount of gas available for
star formation. Stellar feedback heats and disperses the cold and
dense gas after a star formation event. 
In a single star formation event, a stellar particle of $\sim$ 90
$\Msun$ is created. This roughly means the formation of a single
high-mass star embedded in a small stellar cluster. Due to the
resolution limit, our simulation can not follow the details of the
star formation process bellow $\sim$ 10 pc scales, only the overall
net effect. This effect is the formation of a small stellar cluster
with an efficiency of 5\%. As we pointed in \S2.1, the star formation
efficiency of Galactic stellar clusters is high, regardless the details
of their formation. However, although the star formation
efficiency is high, subsequent feedback processes produce a low
average star formation.

\subsection{Volume filling factors in the ISM}

Figure \ref{fig:3} also shows that the net effect of stellar feedback
is to produce the hot phase of the ISM. After the initial strong burst
of star formation, This phase can cover up to 80\% of the total
volume. This represents almost the entire volume above a height of 400
pc from the galactic plane. However, pockets of warm gas are embedded
in this hot flow even at 2 Kpc away from the plane. It has the same
inhomogeneous structure seen in the galactic fountain of figure
\ref{fig:2}. After 100 Myr, $\sim 25$\% of the gas is able to scape
the computational volume.

Most of the hot gas is lost or cooled down after 100 Myr. As a result,
the volume of hot gas decreases because the star formation is low and
the injection of energy is lower than in the initial burst.
The simulation settles into a more quiescent regime in which the volume
occupied by the warm and hot phases oscillates in a self-regulated gas
cycle. In this cycle, bursts of star formation (much smaller than the
initial one) produce super-bubbles and galactic chimneys of hot
gas. Therefore, the volume of hot gas increases. As the star formation
fades, the bubbles cool down and the fraction of hot gas decreases
until the next stellar burst. This pattern reflects the star formation
history. The particular fraction of hot and warm phases at any moment
does depend to the particular star formation history of 10-40 Myr
before that moment.


\subsection{Late stages of evolution}

\begin{figure}[tb!]
\epsscale{1.0}
\plotone{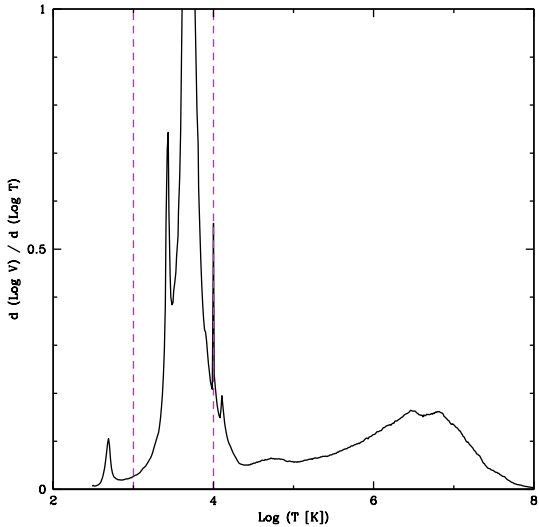}
\caption{Distribution of the gas temperature at 113 Myr. The
distribution has three different peaks corresponding to three
different gas phases of the ISM: cold, warm and hot. Two vertical
lines show the temperature cuts used throughout the paper at $T = 10^3$
K and $T = 10^4$ K. }
\label{fig:5}
\end{figure}

\begin{figure}[tb!]
\epsscale{1.0}
\plotone{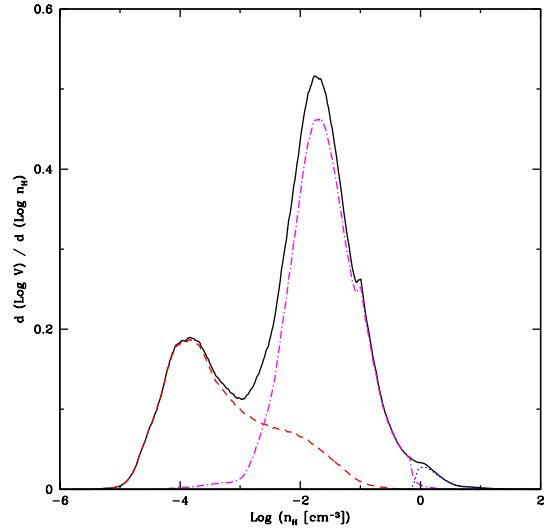}
\caption{Density distribution at 113 Myr and the contribution of the
three phases. The dotted curve shows the cold phase ($T < 10^3$ K),
the dashed curve shows the hot phase ($T >10^4$ K), and the
dash-dotted curve shows the warm phase ($10^3 \thinspace {\rm K} < T
<10^4$ K). The distribution is clearly bimodal. The peaks correspond
to the hot and warm phases. The cold phase dominates the high-density
tail.}
\label{fig:6}
\end{figure}

The latter stages of the simulation offer a more representative view
of the ISM. The effect of the initial conditions is gone. Therefore,
we can study the characteristics of this feedback-driven ISM. We
select a snapshot at 113 Myr, after the second burst of star
formation. At that moment, the warm phase covers $\sim 60\%$ of the
volume and the hot phase filled $\sim 40 \%.$ X-ray emitting gas with
temperatures above $10^{5.5}$ K occupies $\sim 20 \%$ of the volume
inside a height of 250 pc above the galactic plane. This is roughly
consistent with ISM simulations with a 1-pc resolution
\citep{deAvillez04}, Galactic ISM models and observations
\citep{Ferriere98}.

Figure \ref{fig:4} shows representative slices of the box. The medium
is very inhomogeneous at different scales. Large bubbles of low
density coexist with long filamentary structures of dense
clouds. Overall, the medium covers more than 6 orders of magnitude in
density and temperature.  The cold phase forms dense and cold clouds
near the galactic plane. The warm phase fills old cooled bubbles and
low-density clouds. Finally, the hot phase is present in form of hot
bubbles of few hundred pc wide and Kpc-scale chimneys. The gas in
these chimneys is flowing away from the plane with velocities
exceeding $ \pm 300$ km s$^{-1}.$ These bubbles even break the dense
plane in hot spots surrounded by cold and dense shells. All this
phenomenology associated with the hot phase is driven by stellar
feedback. As a result, one effect of the stellar feedback is to
sustain a three-phase ISM.


The distribution of temperature clearly shows the three main peaks of
the three phases of the ISM (Figure \ref{fig:5}). The two local minima
correspond to thermally unstable gas. The minimum around $10^3$ K,
between the peaks of the cold and warm phases, is produced by the
competition of UV heating and radiative cooling. This corresponds to
the unstable regime of the classical two-phase model of the ISM in
thermal equilibrium (Cox 2005). The dip between $10^4$ and $10^5$ K
results from the peak of the cooling curve. The gas cools very fast at
these temperatures. As a result, it usually appears at the interface
between hot and warm gas. As an exception, old bubbles at these
temperatures are present in the simulation with very low densities and
far away from the plane. So, their cooling time is very long. This
temperature distribution supports the temperature cuts used throughout
the paper to distinguish the three phases: $10^3$ K for the cut
between cold and warm phases and $10^4$ for the warm-hot cut. To
summarize, this model of the ISM reproduces the main properties of the
temperature distribution of the ISM (Cox 2005) and predicts that gas
with very high temperatures $10^7-10^8$~K exists in the cores of
galactic chimneys. This gas occupies only 5 \% of the total volume and
have a very small surface density of 4 \timesM 10$^{-6} \Msun$ pc$^{-2}.$

\begin{figure}[tb!]
\epsscale{1.0}
 \plotone{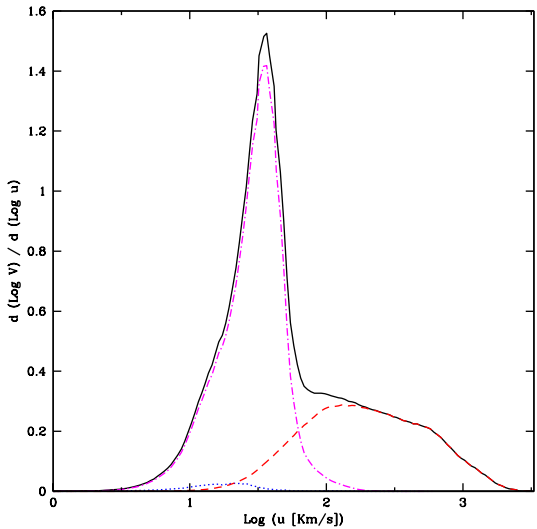}
\caption{Distribution of gas velocity at
113 Myr. The curves represents the 3 gas phases as in figure \ref{fig:6}.
Cold and warm phases have moderate velocities, mostly bellow
100 km s$^{-1}.$ The hot phase dominates the high velocity tail of the
distribution with velocities up to 2000 km s$^{-1}.$ These are
outflows of gas which escapes the system.  }
\label{fig:7}
\end{figure}

\begin{figure}[tb!]
\epsscale{1.0}
 \plotone{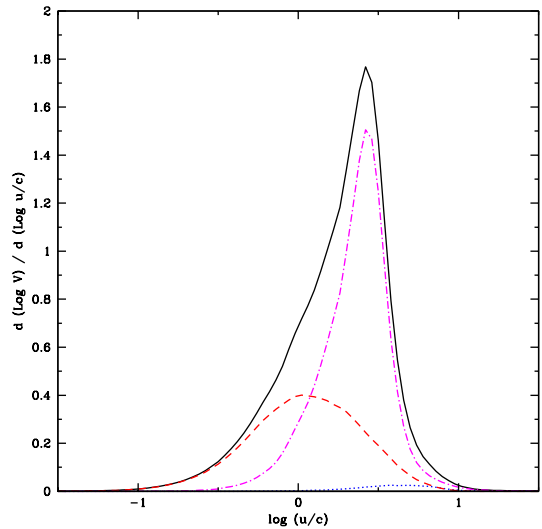}
\caption{ Distribution of the Mach number (u/c).  The curves
represents the 3 gas phases as in figure \ref{fig:6}.  80\% of the
gas has supersonic motions. half of the hot phase has subsonic motions
}
\label{fig:8}
\end{figure}

These three phases of the ISM are also clearly visible in the density
distribution (Figure \ref{fig:6}). We use the temperature cuts defined
before to see the contribution of the different phases. Thus, the hot
phase dominates the low-density range, bellow $10^{-3}$ $\cm.$ The
warm phase covers intermediate densities, and the cold phase dominates
the high density tail above 1 $\cm.$ The density distribution of any
of the three phases cannot be described by a single lognormal
distribution, as claimed in Wada \& Norman (2001). Instead, a
combination of several lognormal distributions may give a better
approximation \citep{RobertsonKravtsov07}.


The distribution of velocities (Figure \ref{fig:7}) shows two distinct
features. The warm phase contribute to a strong peak around 30 km
s$^{-1}.$ The hot phase dominates the high velocity tail. It has
velocities as higher as 2000 km s$^{-1}.$ The gas with velocities in
this tail can easily escape the system. This gas forms hot outflows
and galactic chimneys.


Finally, figure \ref{fig:8} shows the distribution of the Mach number,
\cal{M}=u/c, where u is the gas velocity and c is the sound
speed. The distribution shows that 80\% of the volume has supersonic
motions. Almost all the warm phase, half of the hot phase and all the
cold phase are supersonic flows. The subsonic range is dominated by
the hot phase, In conclusion, the ISM can hold high supersonic
motions, driven by stellar feedback.

\subsection{Degrading resolution}

\begin{figure}[tb!]
\epsscale{1.0}
\plotone{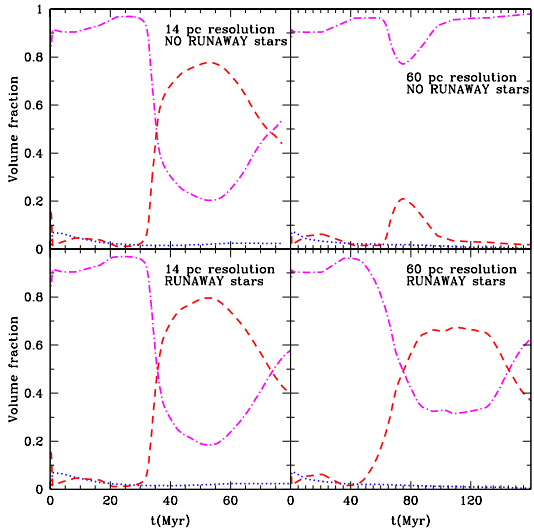}
\caption{The panels show the evolution of the volume occupied by each
phase of the gas in four different models. The curves represents the 3
gas phases as in figure \ref{fig:3}. The hot phase is almost lost for
the low resolution run without runaway stars (top right panel). If
runaway stars are included, the hot phase is recovered at low
resolution. As a result, a fraction of runaway stars produces an
effect on the global ISM and its more evident in low resolution runs.}
\label{fig:9}
\end{figure}

The resolution in cosmological simulations of galaxy formation is much
lower than the simulations of the ISM presented before. So, we can
wonder if this picture of stellar feedback can hold if the resolution
is degraded.  Therefore, the same ISM models have been performed with
high resolution (14 pc) and with low resolution (60 pc).  The fraction
of volume filled with each gas phase is used as a proxy to check the
global effect of stellar feedback in the ISM (Figure \ref{fig:9}).  Left panels
show that the hot phase covers a significant volume. 

At low resolution and without runaway stars (top right panel), the hot
gas is almost absent from the simulation. Small filaments are not resolved
and the subsequent star formation is suppressed in these areas which
can be easily broken by stellar feedback. As a result, star formation
is concentrated at the center of big clumps of gas. Stars inject
energy in high density regions, so this energy is radiated away
without any thermodynamical effect in the medium. 

However, if the model of runaway stars is included, the hot phase is
recovered at low resolution. Stars can now migrate away from high
density regions, so the injection of energy is more efficient in
forming hot gas. As a result, the model with runaway stars can
reproduce the effect of stellar feedback even at a resolution of 60 pc.

\subsection{The expansion of a hot bubble}

\begin{figure}[tb!]
\epsscale{2.0}
\plottwo{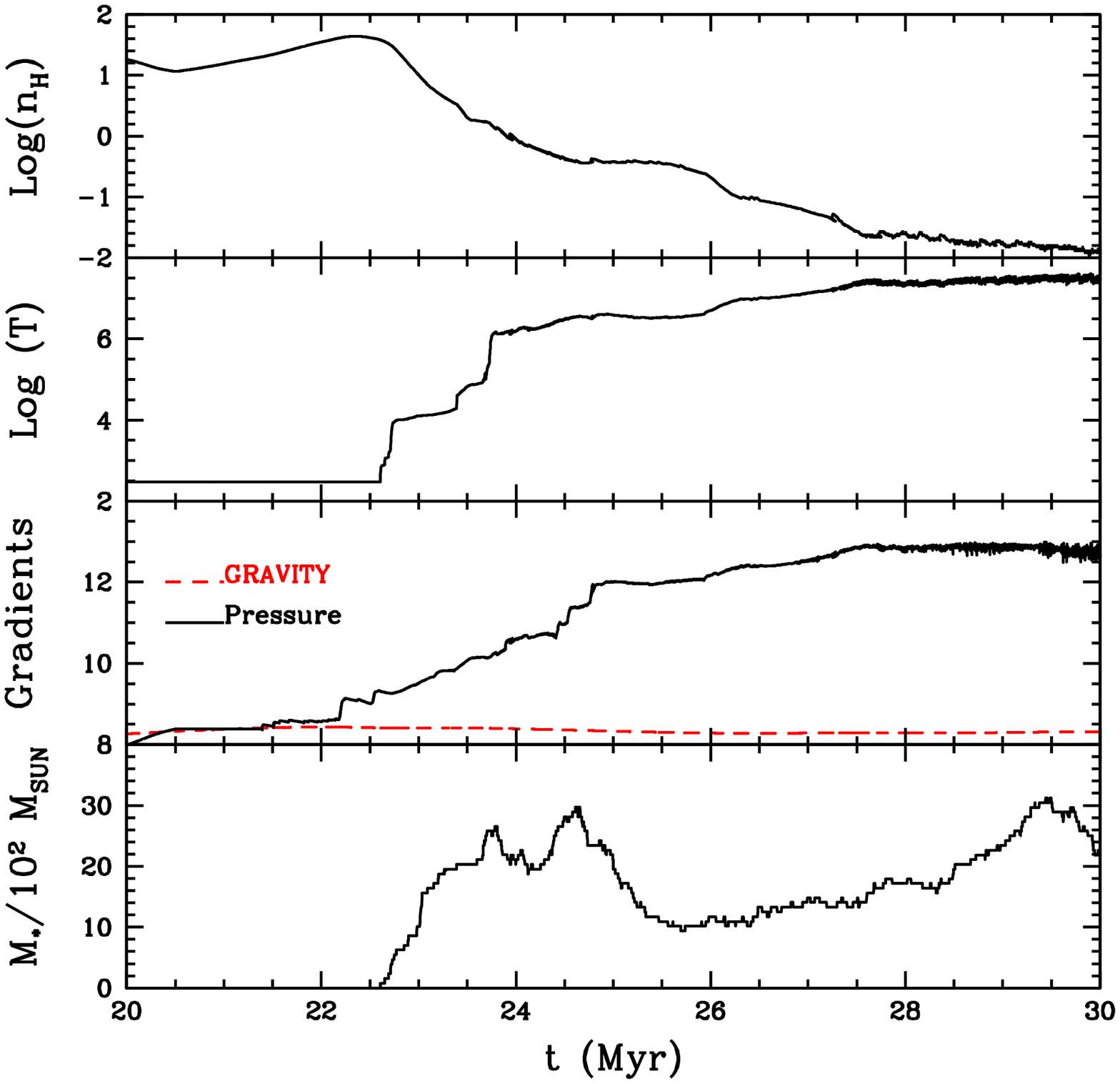}{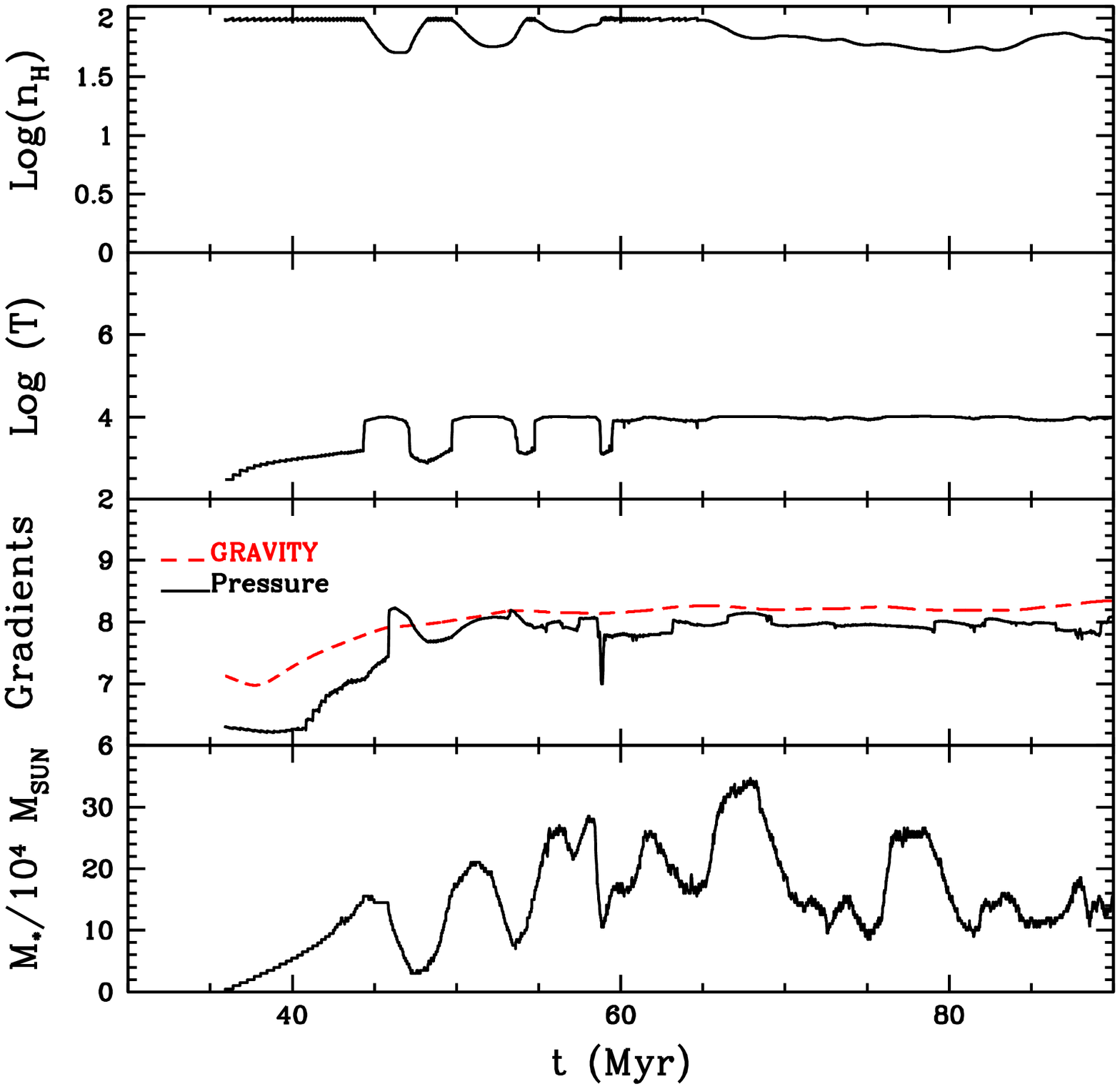}
\caption{Evolution of the properties of a single volume element
(Temperature, density, gradients of pressure and gravity, and the mass
in young stars) in a run with 8 pc resolution (top panel) and with a
resolution of 30 pc (bottom panel). The top panel shows an
over-pressured volume that expands due to stellar feedback. It produces
a hot cavity filled with low-density gas. The bottom panel shows how this
hot bubble cannot develop when the gradients of pressure do not
dominate over gravity.}
\label{fig:10}
\end{figure}

As a example of the conditions of the overheating regime discussed in
section 2, we can ask now how a hot bubble develops in the first
place. The top panel of figure \ref{fig:10} shows the physical
conditions of a single volume element over 10 Myr. This volume
develops a hot and dilute medium starting from a cold and dense phase.
The gradients are computed using a 3-points finite differences
expression using the adjacent cells.

At the beginning, there are no stars present inside that volume. So,
there is no feedback heating. At the same time, the density is high
enough so the the radiative cooling dominates over the UV background
heating. As a result, the medium stays at the floor temperature of 300
K. The medium is also in hydrostatic equilibrium.

The situation drastically changes when young stars appear. They are
not born inside that particular volume. Instead, they are drifting
slowly from adjacent cells. The result is that this young population
injects energy into the medium. So, heating dominates over cooling
initially. The system responds by increasing the temperature. As a
result, the cooling rate increases and the medium reaches a balance
between cooling and heating rates in a very short time-scale. This is
because the cooling time is very short in those conditions.  The net
result is a medium slightly hotter than surroundings so this
over-pressured region expands and the density inside that volume
decreases.

Around $10^4$ K, the cooling curve is a very steep function of
temperature, so the temperature increases very slowly. But, at the
same time, the density drops faster. As a result, the cooling rate
decreases. This expansion is fueled by a roughly constant injection of
energy from massive stars.

When the conditions of eq. \ref{eq:4b} are fulfilled, the medium can
pass through the peak of the cooling curve, somewhere between
$10^4-10^5$ K. After that, the gas has low density and a temperature
of few millions degrees. As a result, heating dominates over cooling
and a hot cavity is formed.

\begin{figure}[tb!]
\epsscale{2.0}
\plottwo{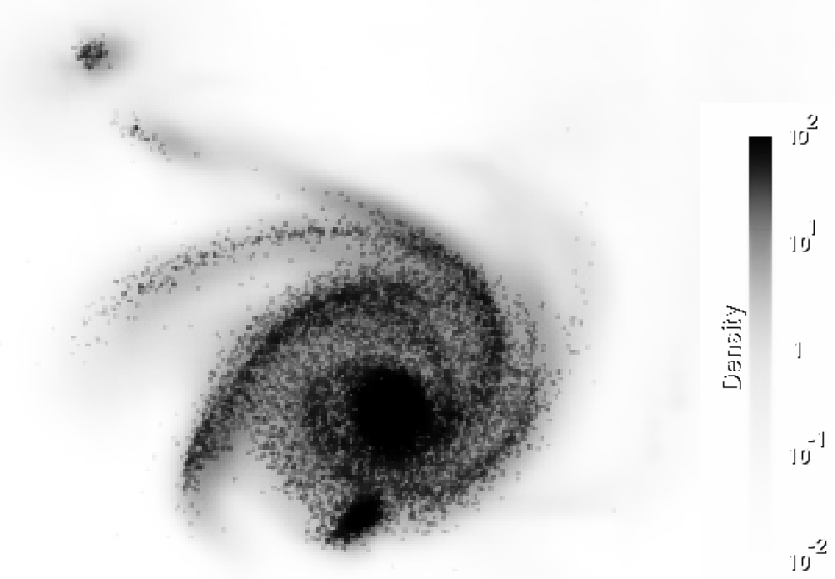}{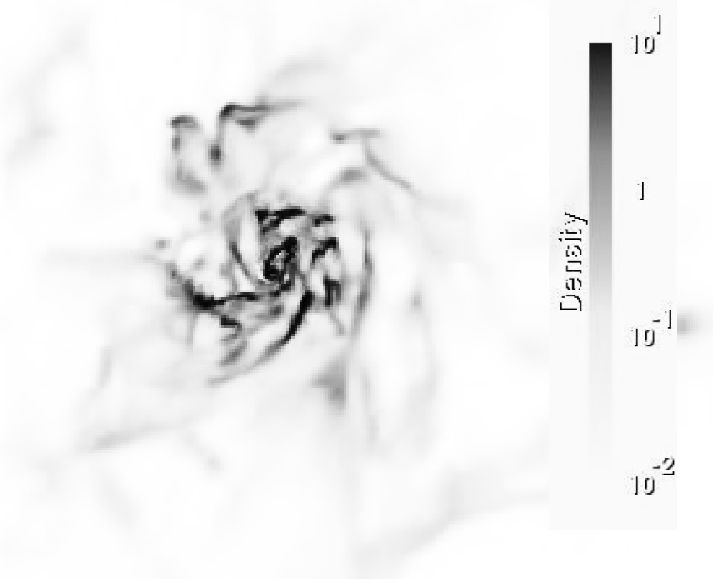}
\caption{Density-weighted average gas density along the line-of-sight
through a MW progenitor at redshift 3. The top panel shows a smooth
gas distribution with some density enhancement due to a pattern of
spiral arms.  Young stars appear as points. They follow the spiral
pattern. However, the distribution has a very concentrated and dense
center. The bottom panel shows the heating regime case. The
distribution is not as dense, so the range of density is
different. The center is less concentrated and the distribution looks
clumpy with dense filaments and clouds embedded in a more diffuse
medium. This is the case of a multi-phase ISM. The size of the images
is 30~Kpc in both cases.}
\label{fig:11}
\end{figure}

The bottom panel of figure \ref{fig:10} shows a different
situation. The volume is selected to be the high-density core of a
molecular cloud formed in the low-resolution run shown in the
top-right panel of figure \ref{fig:9}.
Stellar feedback from the stars formed in that core are able to heat
the gas only to $10^4$ K. The gradients of pressure do not overcome
gravity. The condition of bubble expansion is not fulfilled ,
eq. (\ref{eq:5b}). As a result, the density remains high and a hot
bubble can not develop.

\section{Results of Cosmological runs}

In previous sections, we have shown that our models of stellar
feedback follow the effect of supernova explosions and stellar winds
in the ISM with a resolution of about 50 pc. The result is the
formation of super-bubbles and galactic chimneys. Both are filled with
hot and dilute gas. The net result is a multi-phase ISM and galactic
outflows with large velocities.

Now, we can study the effect of stellar feedback in galaxy
formation. We apply these feedback models in cosmological
hydrodynamics simulations with a similar resolution of 35-70 pc. The
simulations follows the formation of a MW-type galaxy starting from
primordial density fluctuations.

The computational box is 10 $h^{-1}$ Mpc commoving box. We apply a
zooming technique \citep{Klypin01} to select a lagrangian volume
of 3 virial radius centered in a MW-size halo at redshift 0. Then, we
resimulate that volume with higher resolution. The region has a radius
of about 1.5 $h^{-1}$ comoving Mpc. The simulation has about 5 million
dark matter particles. They have three different masses. The
high-resolution region is resolved with 3.4 million dark matter
particles with a $7.5\times 10^5$~\msun~ mass per particle. The
high-resolution volume is resolved with about 17 million volume
elements at different levels of resolution. The maximum resolution is
always between 35 and 70 pc. A short summary of the details of the
simulations is given in table \ref{tab:1}. The cosmological model
assumed throughout the paper has $\Omega_m=0.3,$ $\Omega_\Lambda =
0.7,$ h=0.7 and $\sigma_8=0.9.$


\subsection{Heating regime versus overcooling regime}

\begin{figure}[tb!]
\epsscale{1.0}
\plotone{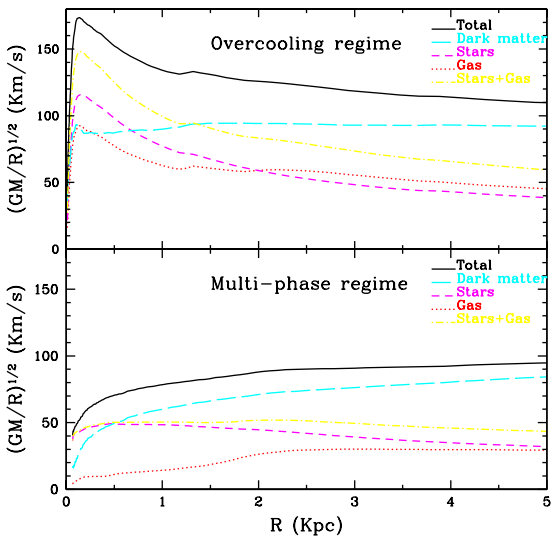}
\caption{Circular velocity profile of the main progenitor of a MW-type
galaxy at redshift 5. The top panel shows the results of an
inefficient stellar feedback. The galaxy is too concentrated and has a
too massive spheroidal component. By contrast, the bottom panel shows
a regime in which stellar feedback is more efficient and it can
regulate the growth of the galaxy. The maximum resolution in both
cases is 70 pc.}
\label{fig:12}
\end{figure}

\begin{figure}[tb!]
\epsscale{1.0}
\plotone{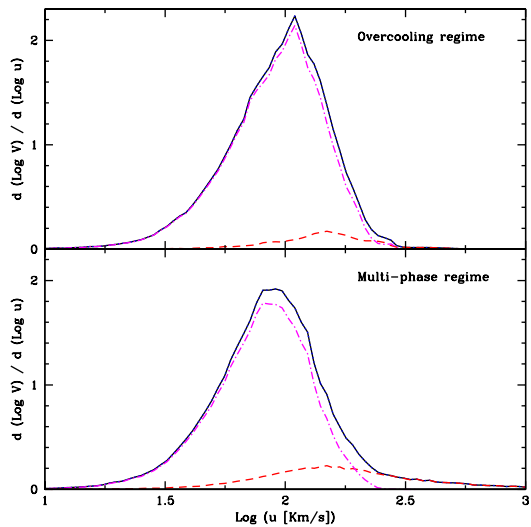}
\caption{Distribution of velocities at the highest levels of the
resolution for two models. The dashed curve represents the hot phase
with $T > 10^4$ K and the dash-dotted curve represents gas with
temperatures bellow $10^4$ K. The bottom panel shows a longer tail of
high-velocities. These are galactic outflows which can reach
velocities exceeding $10^3$ km s$^{-1}.$}
\label{fig:13}
\end{figure}

We compare two cosmological simulations with the same spatial
resolution but different regimes.  Table~\ref{tab:1} shows a summary
of the two simulations. The  over-cooling model has
low star formation efficiency. In addition, the cosmological UV
background according to \citet{Haardt96} is present over the
whole evolution. Finally, a constant model of stellar feedback is
used.

In the second simulation the heating regime develops. It has a high
efficiency of star formation and the UV background is limited to its
value at redshift 8. The supernova model of stellar feedback was used
in this case. We use a model of star formation in which each star
formation event was treated as a random process (see Appendix). In
this way, we keep a moderate galactic star formation rate of $\sim$ 10
$\Msun$ yr$^{-1}$ inside the main galaxy at redshift 3. 
\begin{deluxetable}{lcc} 
\tablecolumns{3}
\tablewidth{0pc} 
\tablecaption{Parameters of cosmological models.\label{tab:1}}
 \tablehead{  \colhead{Parameter}   &  \multicolumn{2}{c}{Models} }
\startdata
Comoving box size & \multicolumn{2}{c}{14.28~Mpc} \\
Number of DM particles & \multicolumn{2}{c}{$5.4\times 10^6$} \\
DM particle mass  &  \multicolumn{2}{c}{$7.5\times 10^5M_{\odot}$} \\
Number of cells  &  \multicolumn{2}{c}{$17.5\times 10^6$} \\
Max. resolution (proper)  &  \multicolumn{2}{c}{35-70~pc} \\
Max. number of stars  &  \multicolumn{2}{c}{$3.7\times 10^6$} \\
Min. mass  of stellar particle  &  \multicolumn{2}{c}{$10^4\Msun$} \\
Model name & Overcooling & Multi-phase \\
UV flux  & H\& M96 & H\& M96 but constant after $z=8$ \\
Star formation time scale $\tau$& $4\times 10^7$~yrs &  $4\times 10^6$~yrs \\
Model for stellar energy release & Constant & SN model \\
Runaway stars & not included & included \\ 
\enddata
\end{deluxetable}


The simulation in the cooling regime has a cold galactic ISM with
temperature close to $10^4$~K. The simulation in the heating regime
develops a 3-phase ISM.  Hot bubbles develop naturally. They produce
galactic chimneys that combine in a galactic wind. As a result,
galactic winds are the natural outcome from stellar feedback.

Figure~\ref{fig:11} shows the main MW progenitor at redshift 3 for both
simulations. The cooling model has a smooth density distribution with
a small enhancement due to a pattern of spiral arms. In contrast, the
multi-phase model develops a clumpy medium of dense clouds
surrounded by low-density bubbles. This is a multi-phase medium.


\begin{figure*}
\epsscale{1.4}
\plotone{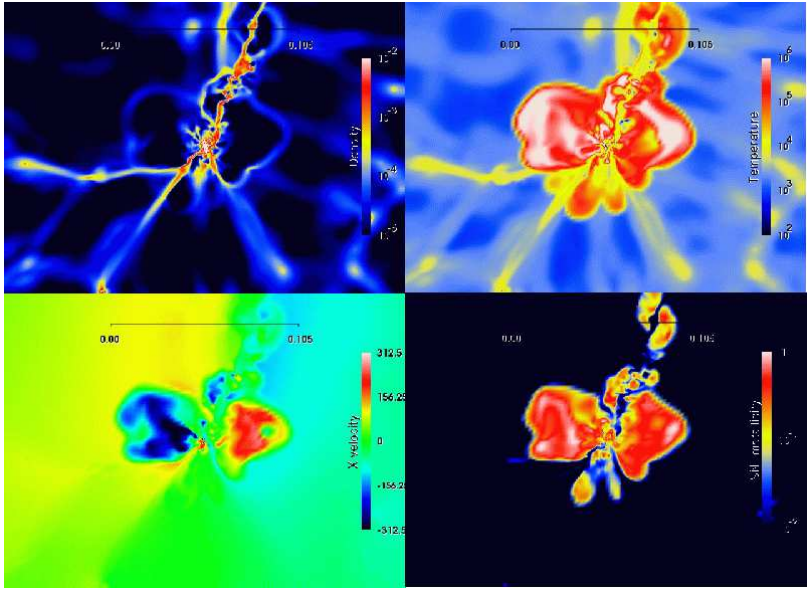}
\caption{This panel shows a galaxy at redshift 3.4 with a resolution
of 45 pc. The figures show slices of 600 Kpc on a side of gas density
(top left), temperature (top right), velocity in the horizontal
direction (bottom left), and metallicity (bottom right). There are
inflows of low-metallicity gas in cold filaments, as well as outflows
of hot, metal-rich gas produced by chimneys in a multi-phase
interstellar medium. Outflow velocities exceeds 300 km s$^{-1}$. The
virial radius is 70 Kpc and the total virial mass is 10$^{11}$ solar
masses.}
\label{fig:14}
\end{figure*}

\begin{figure*}
\epsscale{1.4}
\plotone{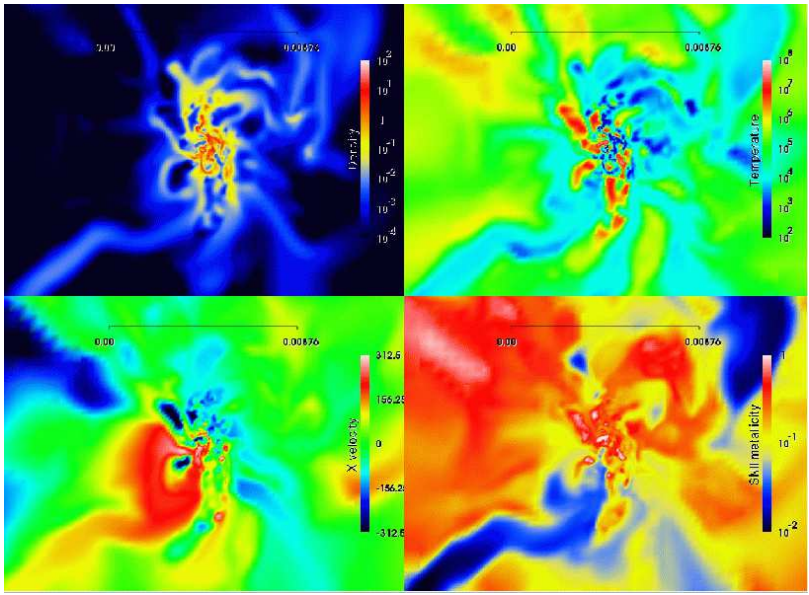}
\caption{The same as in figure \ref{fig:14}, but now the size of the images is 50
Kpc. It shows a multi-phase ISM of cold and dense clouds surrounded by
bubbles of hot and dilute gas. Inflow and outflows velocities can
reach 300 km s$^{-1}$. The outflows are galactic chimneys powered by
core-collapse supernova. Therefore, they are rich in
$\alpha$-elements. In contrast, the inflow of gas has almost
primordial composition.}
\label{fig:15}
\end{figure*}

\subsection{Comparison of circular velocity profiles}

We can see now the effect of this multiphase medium in the galaxy
assembly. We use the profile of circular velocity as a proxy of the
mass distribution, $V_c=\sqrt{GM/R},$ where $G$ is the gravitational
constant and $M$ is the mass inside a radius $R.$ Figure \ref{fig:12}
shows the profile of the circular velocity for the same galaxy in the
two cases. The simulation with the overcooling problem shows a strong
peak in the baryonic component of the circular velocity. Both gas and
stars are very concentrated in the first Kpc.  In contrast, the
simulation with multiphase medium has a more shallow circular velocity
profile. This indicates a less concentrated galaxy with less
baryons. At the virial radius, R$_{vir}=$16 Kpc, the virial mass at
z=5 is 3.1 \timesM 10$^{10}$ $\Msun.$ A large fraction of this mass is
dark matter, M$_{dm}=2.7$ \timesM 10$^{10}$ $\Msun.$ The mass in gas
is M$_{gas}=0.24$ \timesM 10$^{10}$ $\Msun$ and the baryons locked
into stars accounts for only 0.14 \timesM 10$^{10}$ $\Msun.$


\subsection {Galactic winds and multiphase medium}

The hot bubbles in the multi-phase medium develop galactic fountains
that produce hot outflows with very high velocities: larger than
$10^3$ km s$^{-1}.$ These outflows are not produced in the cooling
model. The Figure \ref{fig:13} shows the difference in the
distribution function of velocities for both cases. We take all cells
at the highest levels of refinement. Therefore, we select a volume
close to the galaxies in the simulations. The multi-phase model has a
bigger fraction of hot gas with much larger velocities than in the
cooling model. These outflows contribute to the high-velocity tail of
the distribution. In the cooling model, the distribution drops at 300
km s$^{-1}$, while in the hot case the tail extends beyond $10^3$ km
s$^{-1}.$

These galactic-scale outflows can be seen figure \ref{fig:14}. It
 shows a slice of the simulation through the main MW-progenitor at
 redshift $ z=3.4$. At that redshift, its virial radius is 70~Kpc and
 the total virial mass is 10$^{11}$~\msun. The gas density panel shows
 the galaxy embedded in a cosmological web of filaments. The galaxy at
 the center is blowing a galactic wind of hot and dilute gas with
 outflows velocities exceeding 300 km s$^{-1}.$ The wind is rich in
 $\alpha$-elements and other products of the ejecta of core-collapse
 supernova. These metal-rich outflows can contribute to the enrichment
 of the halo and the inter-galactic medium. These outflows can reach
 even higher velocities and can escape the galactic halo and enrich
 the inter-galactic medium.  The galactic wind is produced by the
 combination of different galactic chimneys anchored in the
 multi-phase ISM of the galaxy.

Figure \ref{fig:15} shows this multi-phase ISM. Cold and dense clouds
coexist with low-density bubbles filled with very hot gas. Warm gas
with intermediate densities and temperatures filled areas of low star
formation and inflows of gas with almost primordial composition.

\subsection{Density profile consistent with a cuspy profile}

\begin{figure}[tb!]
\epsscale{1.0}
\plotone{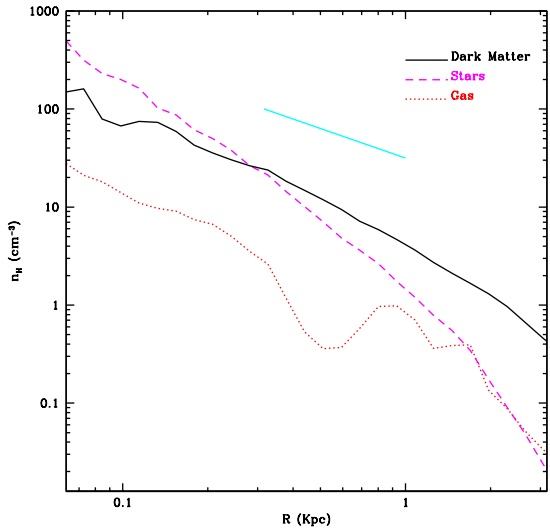}
\caption{Density profile of a galactic halo at redshift 5. The dark
matter distribution is consistent with a cusp. The diagonal line shows
a slope of -1.}
\label{fig:16}
\end{figure}

Figure \ref{fig:16} shows the inner profile of density of the
different components of the galaxy: dark matter, gas and stars at
redshift 5. The density slope of the dark matter profile is consistent
with a cuspy profile.  In contrast, \citet{Mashchenko07} reported the
formation of a core rather than a cusp in the central $\sim$ 300 pc of
a much smaller galaxy at high redshift ( $\sim$ 10$^9$ $\Msun$ at z=6)
in a SPH cosmological simulation.  In their case, the mechanism that
removes the cusp is gravitational heating from large fluctuations in
the gravitational field.  These fluctuations are produced by bulk
motions of gas clumps driven by stellar feedback \citep{Mashchenko06}.
These motions remove episodically 90\% of the mass from the central
100 pc after each burst of star formation.

However, these gas clumps can be overproduced in simulations if the
local Jeans length is not resolved \citep{Truelove97}. This produces
an artificial gas fragmentation and big clumps of stars.  An excessive
clumpiness can artificially increase the efficiency of this
gravitational heating. In our simulations, we prevent this artificial
fragmentation by the implementation of a pressure floor that increases
the effective Jeans length to the resolution limit
\citep{RobertsonKravtsov07}.  However, a direct comparison between our
results and \citet{Mashchenko07} is difficult because we follow the
formation of a much bigger galaxy ($\sim$ 10$^{10}$ $\Msun$ at z=5),
in which the effect of this gravitational heating driven by stellar
feedback is less important. Therefore, these gravitational heating
driven by stellar feedback can not be ruled out in low-mass and
gas-rich starburst galaxies.


\section{Summary and conclusions}

We study the role of supernova explosions and stellar winds in the
formation of galaxies. Our approach is to model these processes
without the ad-hoc assumptions typically used on stellar
feedback. Unlike many currently used prescriptions, we \emph{do not
stop cooling} in regions where the energy from stellar feedback is
released (Thacker \& Couchman 2000; Brook et al. 2004; Keres et
al. 2005; Governato et al. 2007). Moreover, instead of using a
sub-resolution model of a multi-phase medium ( Springel \& Hernquist
2003, Cox et al. 2006), we \emph{resolve} that multi-phase
medium. This is a more straightforward way to model stellar
feedback. It eliminates many ad-hoc assumptions. This approach also
produces naturally the outcomes usually associated to stellar
feedback: hot bubbles, chimneys and galactic winds.

Feedback heating has an effect in the ISM only when it dominates over
radiative cooling. Section 2 shows the necessary conditions for this
heating regime (eq.(\ref{eq:4b}-\ref{eq:5b})). We find that a model of
cooling bellow $10^4$ K is a key ingredient to fulfill these
conditions. Thus, by resolving the conditions of molecular clouds $( T
\approx 100$ K and $n_H>10$ cm$^{-3}),$ we resolve the conditions, in
which stellar feedback is more efficient in the ISM on galactic
scales.

We perform parsec-resolution simulations of a piece of a galactic disk
in order to see the effects of stellar feedback and to test our
models. When we use a realistic feedback and high resolution, the
system has a low star formation rate and it forms hot super-bubbles of
100-pc scales and Kpc-scale galactic chimneys. We found that the cores
of these chimneys reach temperatures of $10^7-10^8$ K, very low
densities $(n_H < 10^{-4}$ cm$^{-3}),$ and outflow velocities
exceeding $10^3$ km s$^{-1}.$

 Then, we degrade the resolution to see if this picture of multi-phase
ISM holds at a resolution that we can achieve in cosmological
simulations. We found that runaway stars help to spread the effect of
stellar feedback. They usually explode as supernovae in low-density
regions, few 100 pc away from their natal molecular cloud. This is an
effect found in nature (Stone 1991), which enhances the feedback. So,
it should be included in any realistic model of stellar feedback.

Thermal feedback from young stars is able to produce long time-scales
of gas consumption by dissipating the star-forming gas. As a result,
although this gas has high star formation efficiency, subsequent
feedback processes produce a low star formation rate, averaged over
all cold and dense gas. For example, in the simulations of the ISM
described in \S 4, the gas with a density above the density threshold
for star formation can form stars with high efficiency. However, the
average star formation efficiency in the simulated clouds is roughly
2.5\% over a free-fall time-scale (\S 4.3). This is roughly consistent
with estimations of the star formation efficiency in molecular clouds
\citep{Zuckerman74,KrumholzMc05,KrumholzTa07}.

In cosmological simulations (\S 5), We find a moderate galaxy star
formation rate, SFR = 10 $\Msun \yr$ and a significant amount of cold
and dense star-forming gas, M$_{\rm dense \thinspace gas}= 10^9
\thinspace \Msun$ inside a 5 Kpc star-forming disk at redshift
3. These values are consistent with observations of nearby starburst
galaxies. Using the observed relation between the star formation rate
and the amount of star-forming gas of \citet{GaoSolomon04}, the star
formation rate expected for 10$^9$ $\Msun$ of cold and dense gas is 20
$\Msun \yr.$ This is close to the value found in your simulations.
Moreover, the galactic gas consumption time-scale of dense gas,
M$_{\rm dense
\thinspace gas} / {\rm SFR}$ is $\sim 100$ Myr.  This is consistent
with observed values in local starburst galaxies which are usually
used as analogs of star-forming galaxies at high redshift
\citep{Kennicutt98}.

In our simulations, star formation proceeds in a way consistent with
observations of star-forming galaxies \citep{Kennicutt98}. From the
numbers given above, the gas surface density of the star-forming disk
of 5 Kpc radius at redshift 3 is $\Sigma_{\rm gas}=13$ $\Msun$
pc$^{-2}.$ Using the Kennicutt fit for nearby star-forming galaxies
\citep{Kennicutt98}, the expected value for the star formation rate
surface density is $\Sigma_{\rm SFR} = 10^{-2} \thinspace \Msun \yr
\kpcTwo.$ The measured value from the simulations is $\Sigma_{\rm SFR}
= 1.3 \times 10^{-1} \thinspace \Msun \yr \kpcTwo.$ Although this
value is an order of magnitude higher than the expected value from the
fit, it is still within the intrinsic spread found in observations.
As a result, our simulated high-redshift galaxy seems more compact
than the average star-forming galaxy at low-redshift.

Our cosmological simulations with this model of stellar feedback do
not have the overcooling problem. The fraction of cold baryons (stars
and gas with a temperature bellow $10^4$ K) inside the virial radius
at z=5 is 0.6 times the cosmological value ( $f_{cosmo}$=0.15). This
is consistent with galaxy mass models \citep{Klypin02}.  Instead of a
cold disk, we produce a multi-phase ISM with the same features seen in
the simulations of the ISM described in section 4: cold clouds, hot
super-bubbles and galactic chimneys. The angular momentum problem is
also reduced. Instead of a compact object with a strong peak in the
rotation curve, we produce more extended galaxies with nearly flat
rotation curves. Baryons are less concentrated when stellar feedback
plays a role in the formation of galaxies. At the same time, the
density profile of dark matter is still consistent with a cuspy
profile.

In this picture, galactic chimneys powered by stellar feedback combine
into a galactic wind. So, galactic winds appear as the natural outcome
of stellar feedback in starburst galaxies at high redshifts. We found
typical outflow velocities of 300 km s$^{-1}$ with some exceptional
examples of outflows exceeding 1000-2000 km s$^{-1}.$ This is
consistent with observation of outflows at high redshift
\citep{Law07}. From a sample of $\approx 100$ galaxies at redshift
$1.9<z<2.6$, \citet{Steidel07} find a mean outflow velocity of 445 km
s$^{-1}$. Some cases have velocities of 1000 km s$^{-1}$.

This picture is only reproduced if the resolution is high enough to
resolve the physical conditions of densities and temperatures of
molecular clouds. Our cosmological simulations reach a resolution of
35 pc, which is 10 times better than the typical resolution in
previous cosmological simulations (Sommer-Larsen et al. 2003; Abadi
et al. 2003; Robertson et al. 2004; Brook et al. 2004; Okamoto et
al. 2005; Governato et al. 2007).

\section*{Acknowledgments}
We are grateful to A. Kravtsov for providing the hydro code. We are in
debt to N. Gnedin for giving us the wonderful analysis and graphics
package IFRIT.  We thank K. Tassis for very useful discussions.  We
acknowledge support of NSF grants to NMSU.  The computer simulations
presented in this paper were performed at the National Energy Research
Scientific Computing Center (NERSC) of the Lawrence Berkeley National
Laboratory and at the NASA Advanced Supercomputing (NAS) Division of
NASA Ames Research Center.

 \appendix

\section{A model of star formation for scales bellow 100 pc}

A successful model of star formation in simulations should take into
account the spatial resolution.  For example, in typical cosmological
simulations with a resolution of $\sim 1$ Kpc, the star formation is
averaged over a large piece of ISM.  These simulations should have a
star formation model with a low star formation efficiency in order to
reproduce the global efficiencies found in nearby galaxies.
Observations of quiescent galactic disks show long gas consumption
time-scales averaged over a significant piece of a galaxy, $\tau_{\rm
global} = \Sigma_{\rm gas}/ \Sigma_{\rm SFR} \sim 1$ Gyr , where
$\Sigma_{\rm SFR}$ is the star formation rate surface density an
$\Sigma_{\rm gas}$ is the gas surface density \citep{Kennicutt98,
Kennicutt07}. At the same time, for starburst galaxies, the global gas
consumption time-scale is much shorter, $\tau_{\rm global} = 0.1$ Gyr
\citep{ Kennicutt98}.

However, if the resolution is high enough to resolve the regions where
star formation mainly occurs, giant molecular clouds, the star
formation efficiency can be much higher: the time-scales for the
formation of Galactic stellar clusters are around few Myr and 10\% -
40\% of the gas is consumed \citep{GreeneYoung92, Elmegreen00}. As a
result, simulations which can resolve the sites of star formation
should have a high star formation efficiency only in the high-density
regions, where molecular clouds can form \citep{TaskerBryan06}.  In
practice, the maximum resolution that we can afford is between 30-70
pc. This limits the maximum density that our simulations can resolve.
For example, if we consider a typical giant molecular cloud of $10^5$
$\Msun$ \citep{Rosolowsky07}, the mean density averaged over 30-80 pc
scales will be 10-200 $\cm.$ This gives an idea of the typical
densities where star formation occurs our simulations.

In our code, star formation is allowed in a time step, $dt_{\rm SF}$,
which is equal to the time step of the 0-Level of resolution. This
time step is controlled by the Courant condition for hydrodynamics and
in our cosmological simulations, $dt_{\rm SF}=$ 1-2 Myr. During this
period of time, a stellar particle can form only where the density and
temperature reach a given threshold: $\rho_{\rm gas} > \rho_{\rm SF}$
and $T_{\rm gas} < T_{\rm SF}.$ Even in these cold and dense regions,
each star formation event is treated as a random event with a
probability $Pr$ to occur. We roughly approximate the fact that
regions with higher densities have a higher probability to host star
formation events by assuming a simplified formula:
\begin{equation}
Pr  =   \frac{ \rho_{\rm gas}} { 100\rho_{\rm SF}} 
\end{equation}
In this way, the number of stellar particles remains in a value that
is not computational prohibited.
In the formation of a single stellar particle, the star formation rate
is proportional to the gas density \citep{Kravtsov}:
\begin{eqnarray}
\frac{ d \rho_{*, \rm young}}{dt} & = & \frac{\rho_{\rm gas}}{\tau}
\label{eq:sf}
\end{eqnarray}
where $\rho_{*, \rm young}$ is the density of new stars, $\rho_{\rm
gas}$ is the gas density and $\tau$ is a constant star formation
timescale.  The density and temperature thresholds used are $\rho_{\rm
SF}=0.035$ $\Msun$ pc$^{-3}$ $(n_H=1 $ $\cm)$ and $T_{\rm SF}=10^4 $
K. In spite of the fact that we allow star formation starting at $10^4
$ K, in practice the vast majority ($>90\%$) of ``stars'' form at
temperatures below 1000~K and more than half of the stars form bellow
300 K and densities larger than 10 $\cm.$

As described in \S2.1, the ratio $\rho_{*,\rm young}/\rho_{\rm gas}$
should be $\sim$ 0.1-0.5 for typical conditions of dense, star-forming
gas. Only in this case thermal feedback can produce over-pressured hot
bubbles in the sites of star formation (eq. \ref{eq:4b}).  Based on
equation \ref{eq:sf}, this ratio of densities can be expressed as
\begin{eqnarray}
\frac{\rho_{*,\rm young}}{\rho_{\rm gas}} & = & \frac{dt_{\rm SF}}{\tau}
\end{eqnarray}
As a result, thermal feedback is only efficient in dense, cold,
star-forming gas if $dt_{\rm SF}/\tau \sim 0.1-0.5.$ This sets the
value of $\tau$, because $dt_{\rm SF}$ is set by the conditions of
hydrodynamics, as explained before: $dt_{\rm SF}=1-2$ Myr. Therefore,
the value of $\tau$ should be in the range 2-20 Myr, consistent with
the gas consumption time-scales during the formation of Galactic
stellar clusters \citep{GreeneYoung92, Elmegreen00}. However, this
high local efficiency of star formation in high-density regions
produces the observed low global efficiency, $\tau_{\rm global}=0.1-1$
Gyr, as discussed in \S 4.3.

\end{document}